\documentclass[12pt]{article}
\pdfoutput=1

\usepackage[english]{babel}
\usepackage{epsf,amssymb,amsmath,bbold,bbm}
\usepackage{graphicx,color}
\usepackage{braket}

\numberwithin{equation}{section} 

\usepackage[colorlinks=true,       
    linkcolor=blue,          
    citecolor=blue,        
    filecolor=blue,      
    urlcolor=blue,   
    linktoc=page	
     ]{hyperref}

\usepackage{cite}

\makeatletter
\renewcommand\section{\@startsection {section}{1}{\z@}%
                               {-3.5ex \@plus -1ex \@minus -.2ex}
                               {2.3ex \@plus.2ex}%
                               {\normalfont\large\bfseries}}
\renewcommand\subsection{\@startsection{subsection}{2}{\z@}%
                                 {-3.25ex\@plus -1ex \@minus -.2ex}%
                                 {1.5ex \@plus .2ex}%
                                 {\normalfont\bfseries}}
\makeatother

\usepackage{xspace}

\newcommand{\LF}{\left(}
\newcommand{\RF}{\right)}
\newcommand{\LT}{\left[}
\newcommand{\RT}{\right]}
\DeclareMathOperator*{\Tr}{\text{Tr}}

\newcommand{\beq}{\begin{equation}}
\newcommand{\eeq}{\end{equation}}

\usepackage{float}
\usepackage{caption}
\usepackage{subcaption}

\begin{document}

\begin{titlepage}

\begin{center}

{\Large \bf  Entanglement versus entwinement  \\ in symmetric product orbifolds }\\
 %

\vskip 10mm

{\large Vijay Balasubramanian$^{a,b}$, Ben Craps$^{b}$, Tim De Jonckheere$^{b}$,  G{\'a}bor S{\'a}rosi$^{a,b}$\\
\vspace{3mm}
}

\vskip 7mm
$^a$ David Rittenhouse Laboratory, University of Pennsylvania, 209 S.33rd Street, Philadelphia PA, 19104, U.S.A \\
$^b$ Theoretische Natuurkunde, Vrije Universiteit Brussel, and \\ International Solvay Institutes,
Pleinlaan 2, B-1050 Brussels, Belgium \\

\vskip 6mm
{\small\noindent  {\tt vijay@physics.upenn.edu, ben.craps@vub.be, tim.de.jonckheere@vub.be, sarosi@sas.upenn.edu}}

\end{center}
\vfill

\begin{center}
{\bf ABSTRACT}\\
\end{center}
We study the entanglement entropy of gauged internal degrees of freedom in a two dimensional symmetric product orbifold CFT, whose configurations  consist of $N$ strands sewn together into ``long'' strings, with wavefunctions symmetrized under permutations.  In earlier work a related notion of  ``entwinement'' was introduced.  Here we treat this system analogously to a system of $N$ identical particles.   From an algebraic point of view, we point out that the reduced density matrix on $k$ out of $N$ particles is not associated with a subalgebra of operators, but rather with a linear subspace, which we explain is sufficient.      In the orbifold CFT, we compute the entropy of a single strand in states holographically dual in the D1/D5 system to a conical defect geometry or a massless BTZ black hole and find a result identical to entwinement.   We also calculate the entropy of two strands in the state that represents the conical defect; the result differs from entwinement.  In this case, matching entwinement would require finding a gauge-invariant way to impose continuity  across strands.

\vspace{3mm}
\vfill
\end{titlepage}

\tableofcontents

\section{Introduction}
Over the past ten years, entanglement entropy has turned out to be a crucial quantity to organize our way of thinking about quantum field theory. It characterizes the amount of information an observer with access to a subsystem of the degrees of freedom can learn about the complementary subsystem. As such it is a measure of correlations in field theory. For example, an observer who only has access to a spatial region $A$, can infer from the entropy how much he or she can learn about the complementary region $\bar{A}$. Because of its extensive nature, one typically expects the entanglement entropy to scale with the volume of the subsystem. However it turns out that the entanglement entropy in the ground state of local Hamiltonians scales with the area of the subsystem, either strictly or in a logarithmically violated way  \cite{Srednicki:1993im,Holzhey:1994we,Vidal:2002rm,Calabrese:2004eu,Gioev:2006zz}. Area law scaling of entropy has first been identified in the context of black hole physics \cite{Bekenstein:1973ur,Hawking:1974sw} and has later been made more precise in the context of holography \cite{Ryu:2006bv,Hubeny:2007xt}. The vast majority of literature on entanglement entropy in holography and in field theory  focuses on entanglement of a spatial subregion $A$. However, its definition as von Neumann entropy of a reduced density matrix only relies on a bipartite splitting of the Hilbert space. In field theories with multiple degrees of freedom, one is not forced to consider a splitting in terms of spatial subregions but one can consider more general splittings such as a bipartition in momentum space \cite{Balasubramanian:2011wt,Agon:2014uxa} or in terms of the internal degrees of freedom. Especially in holographic field theories it is interesting to study the entanglement between internal degrees of freedom, because investigating their correlations and entanglement might be important for understanding the physics of the dual bulk theory at scales smaller than the AdS radius \cite{Susskind:1998dq,Susskind:1998vk}. For example, in the BFSS matrix model, the holographic spacetime can be described as a bound state of $N$ D0 branes. It is natural to investigate the entanglement between the D0 branes to understand better the emergence of the holographic spacetime \cite{Banks:1996vh,Balasubramanian:1997kd, Polchinski:1999br}.   As another example, consider the D1-D5 brane system of $N_1$ D1 branes and $N_5$ D5 branes on $\mathcal{M}^{4,1}\times S^1\times T^4$. At low enough energies, this system is described by gravity on AdS$_3\times S^3\times T^4$, and has a dual  description as a marginal deformation of the symmetric product orbifold $\LF T^4\RF^{N_1N_5}/S_{N_1N_5}$.   Because of the duality, we expect the entanglement entropy of a subset of degrees of freedom of the orbifold theory to have a representation in the dual gravity theory. Motivated by this, a field theoretic quantity called `entwinement'  \cite{Balasubramanian:2016xho,Balasubramanian:2014sra}  has been defined, which is holographically represented by the lengths of non-minimal geodesics in $2+1$ dimensional asymptotically AdS spacetimes. Similar questions about the entanglement between internal degrees of freedom could be posed in matrix string theory \cite{Dijkgraaf:1997vv}.\\

Most of the known field theories with holographic duals have internal gauge symmetries. This complicates the study of entanglement entropy. Even the study of ordinary spatial entanglement entropy is involved in the presence of gauge symmetry because of the non-factorization of the Hilbert space due to non-local gauge invariant degrees of freedom such as Wilson loops that cross the entangling surface \cite{Ghosh:2015iwa,Soni:2015yga,Casini:2013rba,Donnelly:2011hn,Donnelly:2014gva,Radicevic:2014kqa,VanAcoleyen:2015ccp}. The gauge symmetry further complicates the computation of entanglement entropy of a subset of internal degrees of freedom, because typically these dynamical variables are not gauge invariant. In a symmetric product orbifold of $N$ free bosons for example, the bosons transform under the permutation group $S_N$, so one needs a way to appropriately specify a subset of the $N$ bosons to compute a gauge invariant reduced density matrix. A prototypical example where this issue has been considered is the quantum mechanics of identical particles \cite{PhysRevB.63.085311,PhysRevA.64.022303,PhysRevA.64.042310,ECKERT200288}.

Say that one has a system of $N$ identical particles, labelled by position operators $x_1,\ldots,x_N$. The permutation invariance puts a constraint on states of the physical Hilbert space and as such we could view the $S_N$ group as a gauge symmetry. A system of identical particles contains analogous features to gauge theories. For example, we will argue that the non-factorization of the Hilbert space that is apparent when studying spatial entanglement entropy in gauge theories, is analogous to what happens for entanglement entropy of modes of the identical particles. Also, entanglement between gauged degrees of freedom is similar to entanglement between identical particles. One can study for example the one particle reduced density matrix, also called one body density functional, by simply integrating out $x_2,\ldots,x_N$ in the full density matrix. The one particle reduced density matrix will be permutation invariant because the wavefunction is, even if the measure on $x_2,\ldots,x_N$ is not permutation invariant. As we will explain in this paper, the reduced density matrix obtained in this way does not have support on a subalgebra of operators, but rather on a linear subspace.\footnote{In \cite{Lin:2016fqk} a definition of entwinement in terms of the entropy associated to state-dependent subalgebras was proposed.  For identical particles, we do not need to find a closure under multiplication, i.e. an algebra, in order to talk about entropy.   It suffices to discuss a linear subspace of operators that closes under the adjoint operation. For another proposal on associating entropies intrinsic to subspaces of operators, see \cite{Ghosh:2017gtw}.}

In this paper, we will consider a natural extension of this construction for identical particles to the symmetric product orbifold CFT. Namely, the orbifold CFT describes identical strands (i.e.,  elementary pieces of string) in a collection of multiwound strings. We will study the resulting density matrices and corresponding entanglement entropies in a set of states that are interesting from the point of view of holography and represent ground states of specific twisted sectors. More specifically, we focus our attention on twisted sectors that are holographically modeled by conical defects and massless BTZ black holes in the D1-D5 system. We will explicitly compute the single strand entropy in these states as well as the entropy of two strands in the state that represents the conical defect. As we will argue, the entropy of a single strand is proportional to the length of a (not necessarily minimal) geodesic  in the dual geometry and equals the entwinement studied in  \cite{Balasubramanian:2016xho,Balasubramanian:2014sra}.  However, the entropy of two or more strands does not agree with entwinement. This is because entwinement seems to be related to entanglement entropy of multiple continuously connected strands, rather than just entanglement entropy of multiple strands. We will comment more on this point in sec.~\ref{sec:discussion}.  \\

The plan of the paper is as follows. We first recall the ideas behind entwinement in sec.~\ref{sec:entwinement}. In sec.~\ref{sec:EE_IdenticalParticles} we review the definition and construction of entanglement entropy in a system of identical particles. We end this section with comparing this to the algebraic definition of the entropy in section \ref{sec:EE_subalgebra} and point out that the entropy between identical particles is associated to a linear subspace of operators rather than a subalgebra.  In sec.~\ref{sec:EE_orbifolds} we present the computation of entanglement entropy of strands in symmetric product orbifolds. After a short review on symmetric product orbifolds in sec.~\ref{sec:rev_orbifold}, we show how to compute entanglement entropy of multiple strands generically in sec.~\ref{sec:rhoA_EE_general}. In sec.~\ref{sec:EE_OneStrand} we compute the entropy of a single strand in the conical defect state and the massless BTZ black hole state. The entropy of multiple strands in the conical defect state is presented in sec.~\ref{sec:EE_multiple}. We conclude with a summary and with some comments on the entanglement entropy of continuously connected strands in sec.~\ref{sec:discussion}.

\section{Long geodesics and entwinement}\label{sec:entwinement_review}
\label{sec:entwinement}

Let us briefly review the notion of entwinement, as introduced in \cite{Balasubramanian:2014sra}. An intriguing idea, mainly motivated by the Ryu-Takayanagi formula, is that the bulk spacetime in AdS/CFT can be reconstructed from the pattern of entanglement of the dual boundary state. However, it soon became clear that this program cannot be achieved by relying only on entanglement entropies of spatial subregions. This is because certain asymptotically AdS spacetimes possess so called entanglement shadows, regions which are never probed by Ryu-Takayanagi surfaces. While it is in principle possible to reconstruct the bulk in these regions \cite{Dong:2016eik,Faulkner:2017vdd}, one would like to see whether the connection between geometry and entanglement continues to hold. Since entanglement shadows are typically of the size of the order of the AdS radius, this question is also related to the problem of sub-AdS locality.

Simple examples of geometries with entanglement shadows are conical defect geometries, and BTZ black holes in three bulk dimensions. In the case of these geometries, it is known that the entanglement shadows can be probed by boundary anchored extremal, but non-minimal geodesics. These are simple generalizations of the Ryu-Takayanagi surface, however, their boundary dual description is not known. It is argued in \cite{Balasubramanian:2014sra} that they should be related to some notion of entanglement between internal degrees of freedom. 

Let us briefly recap the argument in the case of long geodesics in a conical defect spacetime. One can think about the conical defect spacetimes as $Z_n$ orbifolds of AdS$_3$, where $Z_n$ acts by discrete rotations, and the fundamental region is a ``pizza slice" of AdS. A spatial interval in the conical defect maps to the union of all the $Z_n$ images of the interval in the covering AdS$_3$ space. In this language, the existence of the entanglement shadow is a consequence of a transition in the Ryu-Takayanagi surface for the collection of $n$ intervals in the covering space, as we increase their size. This way, the surface never gets close to the tip of the cone. On the other hand, if we could consider the entanglement entropy of a single interval in covering space, we could probe the entire spacetime. Because of the $Z_n$ identification, doing this intuitively corresponds to selecting a \textit{fraction} of the degrees of freedom inside a spatial subsystem for the state dual to the conical defect. Since the $Z_n$ symmetry is a gauge symmetry, it is not at all clear how to identify this fraction of degrees of freedom. Nevertheless, a quantity called entwinement is proposed in \cite{Balasubramanian:2014sra} to measure this entanglement. The prescription is to ungauge the $Z_n$ symmetry, then calculate an entanglement entropy, and then symmetrize the result under the $Z_n$ symmetry, in order to get a gauge invariant quantity.

In a follow up work \cite{Balasubramanian:2016xho}, this idea was explored from a field theory point of view, in the context of the D1-D5 CFT, where the boundary description of conical defect spacetimes is well understood. The result is that a CFT definition of entwinement can be given in terms of the replica trick, 
using elementary replica twist operators of the ungauged theory that are symmetrized over gauge orbits in a particular way.
This procedure successfully reproduces the lengths of non-minimal geodesics in the conical defect and the massless BTZ spacetimes.
However, with this replica definition, it is unclear how this quantity is related to conventional notions of entanglement. 

In the present paper, we wish to give a definition of reduced density matrices and entanglement entropies associated to degrees of freedom which are gauged under the discrete gauge symmetry. When the gauge group is the permutation group, the arising singlet constraints on the wave functions are very similar to the ones obeyed by identical particles, in which context reduced density matrices to a subset of particles can be defined. This will be the main motivation for our definitions and we now turn to reviewing how this works. The key difference from the definition of entwinement in \cite{Balasubramanian:2016xho} will be that the symmetrization over gauge orbits happens at the level of the wavefunction, which is a more conventional choice than the symmetrization procedure of \cite{Balasubramanian:2016xho}. This conventional symmetrization does not explicitly rely on continuity of long strings across strands, which will lead to differences in the case of winding intervals.

\section{Entanglement entropy in  a system of identical particles}\label{sec:EE_IdenticalParticles}

\subsection{Entanglement between identical particles}\label{sec:EE_id}

Say we have $N$ identical particles, fermions or bosons. The question we want to ask is how say $k$ particles are entangled with the remaining $N-k$ particles, without specifying which particles we are talking about \cite{PhysRevB.63.085311,PhysRevA.64.022303,PhysRevA.64.042310,ECKERT200288}. Let's say we have a position space wavefunction
\beq
\psi(x_1,...,x_N),
\eeq
satisfying $\psi(x_1,...,x_N)=(\pm 1)^{\pi(\theta)}\psi(x_{\theta(1)},...,x_{\theta(N)})$ for all permutations $\theta\in S_N$. Here $\pi$ is the parity of the permutation and bosons correspond to the upper sign, while fermions correspond to the lower one. In quantum chemistry, the $k$-body density functional is defined as
\beq
\label{eq:densityfunc}
\rho^{(k)}(x_1,...,x_k;x_1',...,x_k')=\int dx_{k+1}...dx_{N}\psi(x_1,...x_k,x_{k+1},...,x_N)\psi^*(x_1',...x_k',x_{k+1},...,x_N),
\eeq
i.e., it is a formal partial trace over $N-k$ of the coordinates. More abstractly, for an $N$ particle state\footnote{We omit here the normalization because it is different for fermions and bosons.}
\beq
|\Psi\rangle = \sum_{i_1,...,i_N}\Psi_{i_1...i_N} a_{i_1}^\dagger...a_{i_N}^\dagger|0\rangle,
\eeq
the $k$-particle reduced density matrix is 
\beq
\label{eq:kparticle}
{\rho^{(k)}_{i_1...i_k}}^{j_1...j_k} \sim \Psi_{i_1...i_k i_{k+1}...i_N}(\Psi^*)^{j_1...j_k i_{k+1}...i_N} \, ,
\eeq
where we have traced over $i_{k+1} \cdots i_N$ and omitted combinatorial factors. These are valid density matrices on the $k$-particle Hilbert space $\mathcal{H}_k$. They are positive semidefinite, Hermitian, and have finite trace that can be normalized to one. Therefore, it is meaningful to talk about von Neumann entropy for them. Note that for example for the one particle reduced density matrix for fermions this entropy is bounded from below by the log of the number of particles, instead of zero. This minimal value is obtained if and only if the $N$ particle state is a single Slater determinant, i.e., it is of the form
\beq
v_1^{i_1}...v_N^{i_N}a_{i_1}^\dagger...a_{i_N}^\dagger|0\rangle
\eeq
with $v_a^{i_a}$ arbitrary vectors in the single particle Hilbert space for $a=1,\ldots,N$. Therefore this minimal entropy corresponds to undistillable statistical correlations. Nevertheless, one can produce states that cannot be written in this form and these contain genuine entanglement, which can be equally diverse as in the case of distinguishable constituents \cite{ECKERT200288,PhysRevA.72.022302,PhysRevA.78.022329,PhysRevA.88.052309,Sarosi:2013tha,PhysRevA.90.052303}.

\subsection{Subalgebras versus subspaces}\label{sec:EE_subalgebra}

It is natural to ask how the density matrix \eqref{eq:kparticle} is related to the usual ways of defining a reduced density matrix, which relies on the requirement that $\rho$ reproduces expectation values of the global state in some reduced set of degrees of freedom. The most basic such definition is for factorizing Hilbert spaces $\mathcal{H}=\mathcal{H}_A \otimes \mathcal{H}_{\bar A}$, in which case we associate a density matrix $\rho_A$, acting on $\mathcal{H}_A$, to the global state $|\Psi\rangle \in \mathcal{H}$ by requiring that
\beq
\text{Tr}_{\mathcal{H}_A}(O_A\rho_A) = \langle \Psi | O_A \otimes I_{\bar A} |\Psi \rangle,
\eeq
for all operators $O_A$ acting on $\mathcal{H}_A$. This is a linear equation for $\rho_A$, whose unique solution is the usual partial trace.


Because the Hilbert space of identical particles does not factorize, we can clearly not apply this definition to motivate \eqref{eq:kparticle}.  The problem of a non-factorizable Hilbert space also arises when studying entanglement between spatial degrees of freedom in a gauge theory. In that case, the difficulty can be evaded by resorting to the algebraic definition of a reduced density matrix \cite{Casini:2013rba}.
 Given a state $|\Psi\rangle \in \mathcal{H}$ and a von Neumann subalgebra $\mathcal{A}_s$ acting on $\mathcal{H}$,  we can define a unique density operator $\mathcal{\rho}_{\mathcal{A}_s} \in \mathcal{A}_s$ 
 by requiring that
\beq
\label{eq:reduced_alg}
\text{Tr}'(O\rho_{\mathcal{A}_s})=\langle \Psi|O|\Psi\rangle, \;\;\;\; \forall O\in \mathcal{A}_s.
\eeq
The $\text{Tr}'$ refers to the trace over a ``small Hilbert space" that forms a representation of $\mathcal{A}_s$, but in general is different from $\mathcal{H}$. It is fixed in a way explained e.g. in \cite{Casini:2013rba}. To understand this, recall that a typical algebra $\mathcal{A}_A$ of operators associated to region $A$ will have a nontrivial center $\mathcal{Z}$ because of  the constraints on the Hilbert space  (see e.g.\, \cite{Casini:2013rba,Radicevic:2014kqa,Harlow:2016vwg}).   For example, by Gauss's law, in a $U(1)$ theory without matter the electric flux operator at a location just inside boundary of a subregion must equal the flux just outside, but the latter must commute with all operators inside the region.
 Because the elements of the center commute, we can pick a basis on the total Hilbert space in which all operators in $\mathcal{Z}$ are diagonal. In this basis, the elements of $\mathcal{A}_A$ and $\mathcal{A}_{\bar A}$ which also commute with $\mathcal{Z}$ are block diagonal.  This implies that, instead of factorizing completely, the total Hilbert space now only factorizes on these blocks, i.e. one has
\begin{equation}
\label{eq:nonfact}
\mathcal{H}=\oplus_\alpha \mathcal{H}^\alpha_A \otimes  \mathcal{H}^\alpha_{\bar A},
\end{equation}
where $\alpha$ labels the invariant subspaces of $\mathcal{A}_A$. (See \cite{Casini:2013rba}; the proof that this is true in finite dimensions is given in, e.g., \cite{Harlow:2016vwg}.)  The physical interpretation of $\alpha$ is roughly the value of the electric flux going through the interface between $A$ and $\bar A$, which is forced to be the same on the two subalgebras because of Gauss's law.   The trace over the small Hilbert space $\text{Tr}'$ in \eqref{eq:reduced_alg} is then over $\oplus_\alpha \mathcal{H}^\alpha_A$, that is, we perform a partial trace block by block.

Unfortunately, for identical particles, the $k$-particle reduced density matrix \eqref{eq:kparticle} cannot obviously be motivated in this algebraic framework. It is naturally associated to the following linear subspace of operators, closed under the adjoint
\beq
\label{eq:kparticleoperators}
\mathcal{A}_{k-\text{particle}} = \text{span}\lbrace a^\dagger_{i_1}...a^\dagger_{i_k} a_{j_1}...a_{j_k}\rbrace.
\eeq
These are the $k$-body operators, but this set does not close under multiplication, so it defines a subspace of operators rather than a subalgebra.  This is clearest if we focus on the space of one-particle operators
\beq
\mathcal{A}_{\text{one particle}} = \text{span}\lbrace a^\dagger_i a_j\rbrace.
\eeq
The algebra constructed from this subspace by multiplying its operators would actually generate the entire algebra that conserves particle number. Despite this difficulty, the density matrix \eqref{eq:kparticle} does satisfy a relation like \eqref{eq:reduced_alg}, namely
\beq
\label{eq:reduced}
\text{Tr}^{(k)}(O\rho^{(k)})=\langle \Psi|O|\Psi\rangle, \;\;\;\; \forall O\in \mathcal{A}_{k-\text{particle}} ,
\eeq
where the trace $\text{Tr}^{(k)}$ is over the $k$-particle Hilbert space. For a $k$-particle operator $O$ we have
\beq
O\equiv {O_{j_1...j_k}}^{i_1...i_k} a_{j_1}^\dagger... a_{j_k}^\dagger a_{i_1}...a_{i_k} \quad \rightarrow \quad \text{Tr}^{(k)}(O\rho^{(k)}) = {O_{j_1...j_k}}^{i_1...i_k}{\rho^{(k)}_{i_1...i_k}}^{j_1...j_k} ,
\eeq
and therefore \eqref{eq:reduced} implies that
\beq
{\rho^{(k)}_{i_1...i_k}}^{j_1...j_k}  =\langle \Psi |a_{j_1}^\dagger... a_{j_k}^\dagger a_{i_1}...a_{i_k}  |\Psi \rangle \propto \Psi_{i_1...i_k i_{k+1}...i_N}(\Psi^*)^{j_1...j_k i_{k+1}...i_N} ,
\eeq
so we recover \eqref{eq:kparticle}, which in a coordinate basis is the same as \eqref{eq:densityfunc}. So this density matrix can be thought of as a reduced density matrix reproducing expectation values for the subspace of operators \eqref{eq:kparticleoperators}, even though these operators do not close into an algebra. 

Finally, for later reference we note that the subspace of operators defined by \eqref{eq:kparticleoperators} can be written in a first quantized language
\beq
\label{eq:kparticleoperators_firstquant}
\mathcal{A}_{k-\text{particle}} = \text{span}\lbrace [V_1\otimes V_2 \otimes ...\otimes  V_k \otimes I \otimes ... \otimes I]_{\rm sym}\rbrace,
\eeq
where $V_1,...,V_k$ are arbitrary operators on the single particle Hilbert space, there are $N$ factors in the above tensor product, and $[\quad ]_{\rm sym}$ means symmetrization of the tensor factors for bosons and antisymmetrization for fermions. This spans the same space of operators as \eqref{eq:kparticleoperators} acting on the $N$ particle Hilbert space when $k\leq N$, but it is clear that it only closes under multiplication when $k=N$. This form of $\mathcal{A}_{k-\text{particle}}$ will be useful for direct comparison with the orbifold CFT.

\section{Entanglement entropy in symmetric product orbifolds}\label{sec:EE_orbifolds}
\subsection{Introduction to symmetric product orbifolds}\label{sec:rev_orbifold}
A symmetric product orbifold CFT is formed by starting from a two dimensional seed conformal field theory with target space $M$. We will collectively denote the fields on $M$ as $X(\sigma,t)$ (a free boson in the simplest example). The orbifold CFT arises by taking $N$ copies of $X$ and demanding that the fields are indistinguishable under permutations. The resulting target space is $M^N/S_N$. The orbifold CFT describes a collection of $N$ indistinguishable free strings. Because of the $S_N$ gauging fields need not have period $2\pi$ but can belong to a twisted sector in which the boundary conditions are
\begin{equation}
X_i(2\pi,t) = X_{h(i)}\LF 0,t\RF,\qquad \forall i=1,\ldots,N
\end{equation}
for an element $h\in S_N$. Twisted sectors are created by the action of a twist operator on the untwisted sector. For example, suppose that the fields with indices $1$ to $m$ are sewn together into a long string of winding number $m$ by 
\begin{equation}
\sigma_{(1\ldots m)}(t\rightarrow -\infty): \quad X_i\LF 2\pi,t\RF = X_{(i+1)} \LF 0,t\RF \qquad \forall i=1,\ldots,m-1
\end{equation}
with $X_m\LF 2\pi,t\RF = X_1\LF 0,t\RF$.  If the twisted sector labeled by $h$ contains
 $k$ long strings, with winding numbers $m_i$ satisfying $\sum_i m_i = N$, then
the corresponding twisted sector vacuum state can be represented by the action of $k$ twist operators
\begin{equation}
\ket{\psi_h} = \sigma_{\LF 1,\ldots,m_1\RF}\sigma_{\LF m_1+1,\ldots, m_1 + m_2\RF}\ldots\sigma_{\LF N-m_k+1,\ldots, N\RF}\ket{0} \, .
\end{equation}
All physical states should be $S_N$ invariant so the twisted sector state really corresponds to the conjugacy class $\LT h\RT$, such that
\begin{align}
\ket{\Psi_{\LT h\RT}} &= \sum\limits_{g \in S_N} \ket{\psi_{ghg^{-1}}},\nonumber\\
&= \sum\limits_{g\in S_N} \sigma_{\LF g(1),\ldots,g(m_1)\RF}\sigma_{\LF g(m_1+1),\ldots, g(m_1 + m_2)\RF}\ldots\sigma_{\LF g(N-m_k+1),\ldots, g(N)\RF}\ket{0}.
\end{align}
Strictly speaking, the state should also be appropriately normalized. In the $X_i$ basis, 
the gauge invariant wavefunction is 
\begin{align}
\Psi_{\LT h\RT}\LF X_1,\ldots,X_N\RF &= \langle X_1,\ldots,X_N| \Psi_{\LT h\RT}\rangle = \sum\limits_{g\in S_N} \psi_{ghg^{-1}} \LF X_1,\ldots,X_N\RF,\\
&=\sum\limits_{g\in S_N} \psi_h\LF X_{g(1)},\ldots,X_{g(N)}\RF.
\end{align}
Note that the wavefunctions $\psi_h$ can be decomposed into a product of wavefunctions of single long strings which we could specify by $\psi_{m_j}$ such that
\begin{equation}
\Psi_{\LT h\RT}\LF X_1,\ldots,X_N\RF = \sum\limits_{g\in S_N} \psi_{m_1}\LF X_{g(1)},\ldots,X_{g(m_1)}\RF\ldots\psi_{m_k}\LF X_{g(N-m_k+1)},\ldots, X_{g(N)}\RF.
\end{equation}
In this paper we will focus on two states which are interesting from the point of view of holography. The first state contains $N/m$ strings of length $m$ whose wavefunction is
\begin{equation}
\Psi_{m}\LF X_1,\ldots,X_N\RF = \sum\limits_{g\in S_N} \prod\limits_{j=1}^{N/m} \psi_{m}\LF X_{g(jm+1)},\ldots, X_{g((j+1)m )}\RF.
\label{eq:con_def_state}
\end{equation}
In the context of the D1-D5 brane system this wavefunction is holographically dual to a $\LF \text{AdS}_3\times S^3\RF/\mathbb{Z}_m\times T^4$ spacetime \cite{Balasubramanian:2000rt,Balasubramanian:2005qu}. We therefore call this a `conical defect state'. A second state of interest contains $N_m$ strings of length $m$ where
\begin{equation}
N_m = \frac{8}{\sinh\LF m\sqrt{\frac{2\pi}{N}}\RF}.
\label{eq:Nm_btz}
\end{equation} 
This state is dual in the D1-D5 system to BTZ$_0\times S^3\times T^4$, where BTZ$_0$ is the massless three dimensional black hole \cite{Balasubramanian:2005qu,Cvetic:1998xh}.

The wavefunctions described above only have support on continuous configurations.
To see this, consider a twisted sector vacuum state. 
Suppose that a factor of a particular term of the gauge invariant wavefunction has support on a discontinuous configuration of string of length $m$.  This factor can be written in terms of the on-shell action for the configuration
\begin{equation}
\psi_m\LF X_{g(1)},\ldots,X_{g(m)}\RF\equiv \psi_m\LF \mathcal{X}(\sigma)\RF\sim e^{-S_{\text{on-shell}}\LT \mathcal{X}\RT},
\end{equation}
because we are discussing a vacuum state constructed from the Euclidean path integral.   
When the seed CFT is that of a single free boson $X(\sigma)$, the on-shell action can be computed by solving the Laplace equation on a disk with boundary conditions $\mathcal{X}(\sigma)$ at Euclidean time $\tau=0$ and from it computing the Euclidean action. The simplest discontinuous boundary profile is a piecewise constant $\mathcal{X}(\sigma)=0$ when $0\leq \sigma \leq \alpha$ and $\mathcal{X}(\sigma)=\mathcal{X}_0\neq 0$ when $\alpha\leq \sigma \leq 2\pi m$.  The Laplace equation is conformally invariant, so the disk with such a boundary condition can be mapped to the infinite strip with $\mathcal{X}=0$ on the lower boundary and $\mathcal{X}=\mathcal{X}_0$ on the upper boundary. Physically, the problem of determining the Euclidean action of a free field on the infinite strip with two constant boundary conditions is the same as determining the potential energy between two capacitor plates each at constant potential. Clearly the energy between the capacitor plates is infinite. The on-shell action thus diverges and the wavefunction correspondingly vanishes. The same conclusion holds for excited states because adding oscillations to a discontinuous configuration will not remove the divergence of the action.

Similar considerations dictate that the overlap between different twisted sectors is zero \cite{Dixon:1985jw,Lunin:2000yv,Pakman:2009zz}. We can argue this by observing that the boundary conditions do not match in the overlap of different twisted sectors.   Physically speaking, the overlap vanishes because, in a free orbifold of the kind we are considering,  the winding strings cannot split and rejoin into a different configuration.

In summary, states of the physical Hilbert space satisfy two important constraints: (1) they are invariant under $S_N$, and (2) their wavefunctions vanish on discontinuous configurations.

\subsection{Reduced density matrix and entanglement entropy}\label{sec:rhoA_EE_general}

Having defined gauge invariant wavefunctions in the symmetric product orbifold CFT, it is easy to define a gauge invariant reduced density matrix. We define it analogously to 
\eqref{eq:densityfunc} in a system of identical particles. In this definition, the reduced density matrix that has support on the full circle on $l-1$ strands and on an interval $A$ of size $\ell$ on one strand  is
\begin{align}
\rho^{(l)}_A \LF X_1, \right. &\left.\ldots,X_{l-1},X_{l,A};X'_1,\ldots,X'_{l-1},X'_{l,A}\RF\label{eq:general_rhoA}
 \\
= 
\int & DY_{l+1}\ldots DY_N DY_{l,\bar{A}}\Psi^{*}_{\LT h\RT}\LF X'_1,\ldots,X'_{l,A},Y_{l,\bar{A}},\ldots,Y_{N}\RF\Psi_{\LT h\RT}\LF X_1,\ldots,X_{l,A},Y_{l,\bar{A}},\ldots,Y_{N}\RF,\nonumber
\end{align}
where $DY_i = \prod\limits_{\sigma=0}^{2\pi} dY_{l+1}(\sigma)$ and $D Y_{i,\bar{A}} = \prod\limits_{\sigma=\ell}^{2\pi} dY_{l+1}(\sigma)$. 
Compared to the identical particle case a subtlety here is that the measure is infinite dimensional. In this definition, therefore, 
the reduced density matrix is defined in terms of 
 functional integrals rather than ordinary integrals. Because the wavefunction is gauge invariant, the reduced density matrix will be gauge invariant as well even if the measure has not been symmetrized over $S_N$. Another way to write the reduced density matrix is
\begin{align}
\rho^{(l)}_A = \Tr\limits_{\ket{Y_{l,\bar{A}},Y_{l+1},\ldots,Y_N}} \LT \ket{\Psi_{\LT h\RT}}\bra{\Psi_{\LT h\RT}}\RT \, .
\end{align}
Because the state is a sum over gauge copies in $S_N$, we get a double sum over $S_N$,
\begin{align}
\rho^{(l)}_A = \sum\limits_{g,\tilde{g}\in S_N}\Tr\limits_{\ket{Y_{l,\bar{A}},Y_{l+1},\ldots,Y_N}} \LT \ket{\psi_{ghg^{-1}}}\bra{\psi_{\tilde{g}h\tilde{g}^{-1}}}\RT.
\end{align}
One of the two sums can be further split into elements  $\tilde{g}=gc$ where $c$ belongs to the centralizer of $\LT h\RT$, i.e., $c\in C_h\subseteq S_N$, and elements $\tilde{g}=gc$ for which $c\notin C_h$. Elements $c\in C_h$ by definition satisfy $ch=hc$ hence in this case $\tilde{g}h\tilde{g}^{-1}=ghg^{-1}$. The reduced density matrix correspondingly takes the form
\begin{align}
\rho^{(l)}_A =&\left|C_h\right|\sum\limits_{g\in S_N}\Tr\limits_{\ket{Y_{l,\bar{A}},Y_{l+1},\ldots,Y_N}} \LT \ket{\psi_{ghg^{-1}}}\bra{\psi_{ghg^{-1}}}\RT \label{eq:rhoA_cinC_h_notinC_h}\\
&+\sum\limits_{g \in S_N}\sum\limits_{c\notin C_h}\Tr\limits_{\ket{Y_{l,\bar{A}},Y_{l+1},\ldots,Y_N}} \LT \ket{\psi_{ghg^{-1}}}\bra{\psi_{g chc^{-1} g^{-1}}}\RT .\nonumber
\end{align}
If we focus on $l=1$ or $l=2$ where one of the strands only has support on an interval of size $\ell<2\pi$, then the sums over $c\notin C_h$ do not contribute, as we now explain.

First consider tracing out everything and look at an overlap $\langle \psi_{ghg^{-1}}\left. \right| \psi_{gchc^{-1}g^{-1}}\rangle$ where $c\notin C_h$. The wavefunction $\psi_{ghg^{-1}}$ only has support on continuous configurations that should satisfy the continuity constraints of the twisted sector $\LT h\RT$ with a specific ordering of the fields specified by $ghg^{-1}$. Similarly the conjugated wavefunction should satisfy continuity constraints of $\LT h\RT$ with an ordering of fields specified by $gchc^{-1}g^{-1}$. Since $c \notin C_h$, the two sets of constraints are different and the configurations in the overlap can only meet both of them on a submanifold of the integration space.  Hence, overlaps of this type vanish.

  When we consider a reduced density matrix instead, the question is whether the difference in continuity constraints between the wavefunction and its conjugate affect the integration variables or not. In the case of $l=1$ or $l=2$ (with one strand having support only on $\ell<2\pi$) there clearly does not exist any permutation $c$ that leaves the integrated variables unaffected.
Therefore, the sum over $c \notin C_h$ will again not contribute.  The reduced density matrix in these cases becomes
\begin{align}
\rho^{(l)}_A =\left|C_h\right|\sum\limits_{g\in S_N}\Tr\limits_{\ket{Y_{l,\bar{A}},Y_{l+1},\ldots,Y_N}} \LT \ket{\psi_{ghg^{-1}}}\bra{\psi_{ghg^{-1}}}\RT.\label{eq:rhoA_cinC_h}
\end{align}
For $l=2$ and $\ell=2\pi$, or for $l>2$ it is possible to have permutations $c$ that do not affect the integrated variables, and these can lead to contributions from the $c\notin C_h$ terms.

From the algebraic perspective, consider the linear subspace of non-twist observables
\beq
\mathcal{A}_l=\text{span}\lbrace [O_1\otimes ... \otimes O_l \otimes I \otimes ... \otimes I]_{\rm sym}\rbrace,\label{eq:subspace}
\eeq
where $O_1,...,O_l$ are (not necessarily local) operators in the seed CFT and $[]_{\rm sym}$ denotes summing over the images over $S_N$.   Because of the symmetrization, the subspace contains operators that act on any subset of $l$ strands, and thus multiplying elements would have generated the entire non-twisted operator algebra. The same thing happens with the subspace of $k$ particle operators \eqref{eq:kparticleoperators_firstquant} in the case of identical particles.  Following  sec.~\ref{sec:EE_subalgebra} we can define a reduced density matrix as the element of the subspace that computes expectation values of operators in the subspace via tracing.   
Since this density matrix must be constructed out of non-twist operators it cannot contain any elements linking strings twisted by different group elements  (even if the group elements in question belong to the same conjugacy class).   This means including just the $c \in C_h$ terms in  \eqref{eq:rhoA_cinC_h_notinC_h}.

For $l>2$, for which the second term in \eqref{eq:rhoA_cinC_h_notinC_h} need not vanish, if one wants to construct the density matrix restricted to the subspace  \eqref{eq:subspace} of non-twist operators, one must project out the $c \notin C_h$ terms. For the cases we treat in this paper, namely $l=1$ and $l=2$ with $\ell<2\pi$, $\rho_A^{(l)}$ automatically belongs to this subspace. 

Once the reduced density matrix has been computed, the von Neumann entropy can be calculated in the usual way via
\begin{equation}
S\LF \rho^{(l)}_A\RF = -\Tr\LF \rho_A^{(l)}\log\rho_A^{(l)}\RF.
\end{equation}

\subsection{Single strand}\label{sec:EE_OneStrand}
In this section we will compute the single strand reduced density matrix on a spatial interval $A$ and compute its corresponding entanglement entropy. We will do so in the conical defect state and in the massless BTZ state. We will show that the single strand entanglement entropy agrees with the entwinement of a single strand \cite{Balasubramanian:2016xho}. 
\subsubsection{Conical defects: reduced density matrix and entanglement entropy}
Because the conical defect state describes $N/m$ strings, all of equal length $m$, every term in \eqref{eq:rhoA_cinC_h} contributes equally. $N/m -1$ of them can be integrated out completely. Elements of $\rho_A^{(1)}$ are therefore computed by

\begin{align}
\rho^{(1)}_A \LF X_{1,A}, X'_{1,A}\RF = \left|S_N\right|\left|C_h\right| \int \limits DY_{1,\bar{A}}\ldots DY_m\psi_m\LF X_{1,A},Y_{1,\bar{A}},\ldots,Y_m\RF\psi^*_m\LF X'_{1,A},Y_{1,\bar{A}},\ldots,Y_m\RF
\end{align}
If we assume that the wavefunctions $\psi_m$ are normalized to unity, the properly normalized density matrix is
\begin{align}
\hat{\rho}^{(1)}_A \LF X_{1,A},X'_{1,A}\RF &\equiv \frac{\rho_A^{(1)}\LF X_{1,A},X'_{1,A}\RF}{\Tr\rho_A^{(1)}},\\
&= \int DY_{1,\bar{A}} \ldots DY_m\psi_m\LF X_{1,A},Y_{1,\bar{A}},\ldots,Y_m\RF\psi^*_m\LF X'_{1,A},Y_{1,\bar{A}},\ldots,Y_m\RF.\nonumber
\end{align}
Remember that $\psi_m$ is the vacuum wavefunction of a CFT on an $m$-wound string on the extended Hilbert space (i.e., without symmetrization). If $\ell$ is the angular extent of the interval $A$, then by a conformal map, the reduced density matrix can be represented as a path integral on the plane with a cut on the unit circle of angular extent $\ell/m$ \cite{Holzhey:1994we,Calabrese:2009qy}. We know by the usual replica trick that such a density matrix has an entanglement entropy equal to
\begin{equation}
S\LF\hat{\rho}^{(1)}_A\RF = -\Tr\LF \hat{\rho}_A^{(1)} \log \hat{\rho}_A^{(1)}\RF = \frac{c}{3}\log\LT \frac{2m}{\epsilon}\sin\LF \frac{\ell}{2m}\RF\RT.\label{eq:EE_con_single}
\end{equation}
As usual the entanglement entropy is UV divergent and is proportional to the central charge.   The central charge that appears here is the seed central charge of the CFT with target space $M$, which makes sense since we have computed the entropy of a single (collective) field $X_1$ on $A$. In contrast, the spatial entanglement entropy of a collection of $N$ fields would be proportional to the central charge of the orbifold theory $c_N=Nc$. Notice that the entwinement studied in \cite{Balasubramanian:2016xho} is also given by \eqref{eq:EE_con_single}.\\

It is interesting to compare this to the holographic spatial entanglement entropy. When specializing to the D1-D5 orbifold CFT, the single strand entanglement entropy agrees with the length of a minimal geodesic in the dual conical defect geometry when $\ell<\pi$. When $\ell>\pi$, then \eqref{eq:EE_con_single} computes the length of a non-minimal but non-winding geodesic in the conical defect background. The spatial entanglement entropy, on the other hand, as computed by the Ryu-Takayanagi formula \cite{Ryu:2006bv}, would agree with the length of a minimal geodesic both when $\ell<\pi$ and $\ell>\pi$. Note that these matches are somewhat surprising because the orbifold CFT is at a different point in the moduli space of the D1-D5 system than the regime where classical supergravity is valid \cite{Seiberg:1999xz,Larsen:1999uk};  perhaps the form of the entanglement entropy and entwinement are sufficiently constrained by conformal symmetry that they remain the same at different points in moduli space.

\subsubsection{Massless BTZ: reduced density matrix and entanglement entropy}

A twisted sector state is characterized by the set $\{N_m\}$ of numbers of $m$-wound strings with $\sum\limits_{m=1}^N mN_m=N$. Suppose we wish to compute the reduced density matrix $\rho_A^{(1)}\LF X_{1,A},X_{1,A}\RF$. Because \eqref{eq:rhoA_cinC_h} involves a sum over $S_N$, the elements $X_{1,A}$ and $X'_{1,A}$ can occur in strings of various lengths $m$. As is clear from the structure of \eqref{eq:rhoA_cinC_h}, both $X_{1,A}$ and $X'_{1,A}$ have to belong to the same strand. For a fixed embedding of $X_{1,A}$ and $X'_{1,A}$ there are $(N-1)! = |S_N|/N$ terms that give the same contribution. On top of that there are $mN_m$ possible embeddings into strings of length $m$. After integrating out all strings to which $X_{1,A}$ and $X'_{1,A}$ do not belong, elements of the one strand reduced density matrix for general $\{N_m\}$ take the form
\begin{align}
&\rho_A^{(1)}\LF X_{1,A},X'_{1,A}\RF =\label{eq:rhoA_btz_single}\\
& \sum\limits_{m=1}^N\frac{mN_m}{N}\left|S_N\right| \left|C_h\right|  \int DY_{1,\bar{A}}\ldots DY_m\psi_m\LF X_{1,A},Y_{1,\bar{A}},\ldots,Y_m\RF\psi^*_m\LF X'_{1,A},Y_{1,\bar{A}},\ldots,Y_m\RF.\nonumber
\end{align}
We will use the notation  $\hat{\chi}^{(1)}_{m,A}$ for the single strand density matrix on an interval $A$ in an $m$-mound string.  After normalization
this becomes
\begin{equation}
\hat{\rho}_A^{(1)}\LF X_{1,A},X'_{1,A}\RF =  \sum\limits_{m=1}^N\frac{mN_m}{N}\hat{\chi}_{m,A}^{(1)}\LF X_{1,A},X'_{1,A}\RF.\label{eq:reduced_generic}
\end{equation}
The massless black hole typical state has $\{N_m\}$ given by \eqref{eq:Nm_btz}. We will bound the von Neumann entropy of the reduced density matrix on this typical state both from above and from below and show that in the large $N$ limit, both limits converge to the same value. The reduced density matrix \eqref{eq:reduced_generic} is written as a convex combination of density matrices. By concavity of von Neumann entropy, there is a simple lower bound namely
\begin{equation}
S\LF \hat{\rho}^{(1)}_A\RF\geq \sum\limits_{m=1}^N \frac{mN_m}{N}S\LF \hat{\chi}^{(1)}_{m,A}\RF.
\end{equation}
In \cite{Balasubramanian:2016xho} it was shown that the expression on the right hand side to leading order in the large $N$ limit reduces to the one strand reduced density matrix of a single $\sqrt{N}$-wound string, so
\begin{equation}
S\LF \hat{\rho}^{(1)}_A\RF\geq S\LF\hat{\chi}^{(1)}_{\sqrt{N},A}\RF\approx \frac{c}{3}\log\LF \frac{\ell}{\epsilon}\RF.\label{eq:S_lower}
\end{equation}
To establish the upper bound, we use positivity of the relative entropy $S\LF \hat{\rho}^{(1)}_A\left|\right.\hat{\chi}^{(1)}_{\sqrt{N},A}\RF$. Its definition is
\begin{align}
S\LF \hat{\rho}^{(1)}_A\left|\right.\hat{\chi}^{(1)}_{\sqrt{N},A}\RF &\equiv\Tr\LF \hat{\rho}^{(1)}_A\log \hat{\rho}^{(1)}_A\RF - \Tr\LF  \hat{\rho}^{(1)}_A\log \hat{\chi}^{(1)}_{\sqrt{N},A}\RF,\\
&=-S\LF \hat{\rho}^{(1)}_A\RF + S\LF \hat{\chi}^{(1)}_{\sqrt{N},A}\RF-\langle K_{\chi_{\sqrt{N},A}}\rangle_{\chi_{\sqrt{N},A}}+ \langle K_{\chi_{\sqrt{N},A}}\rangle_{\rho_{A}},
\end{align}
where $K_{\chi_{\sqrt{N},A}}\equiv-\log \hat{\chi}^{(1)}_{\sqrt{N},A}$ is the modular Hamiltonian of $\hat{\chi}^{(1)}_{\sqrt{N},A}$. Positivity of the relative entropy implies that
\begin{equation}
S\LF \hat{\rho}^{(1)}_A\RF - S\LF \hat{\chi}^{(1)}_{\sqrt{N},A}\RF\leq \langle K_{\chi_{\sqrt{N},A}}\rangle_{\rho_{A}} - \langle K_{\chi_{\sqrt{N},A}}\rangle_{\chi_{\sqrt{N},A}} .\label{eq:S_upper}
\end{equation}
$K_{\chi_{\sqrt{N},A}}$ is the vacuum modular Hamiltonian on an arc of length $\ell$ in a circle of size $2\pi\sqrt{N}$. Such a modular Hamiltonian can be written as an integral of the vacuum stress tensor \cite{Casini:2011kv,Wong:2013gua}:
\begin{equation}
K_{\chi_{\sqrt{N},A}} = 4\pi\sqrt{N}\int\limits_0^\ell d\theta \frac{\sin\LF \frac{\ell -\theta}{2\sqrt{N}}\RF\sin\LF \frac{\theta}{2\sqrt{N}}\RF}{\sin\LF \frac{\ell}{2\sqrt{N}}\RF} T_{00}\LF \theta\RF.
\end{equation}
The expectation value of the modular Hamiltonian thus becomes an expectation value of the stress tensor in the reduced density matrix on $A$. The expectation value of any local operator  $O$ should satisfy $\langle O\rangle_{\chi_{A,m}} = \langle O\rangle_{\chi_m}$ where $\chi_m$ is the full state of an $m$-wound string. The vacuum expectation value of the stress tensor on a cylinder of circumference $2\pi m$ is \cite{DiFrancesco:1997nk}
\begin{equation}
\langle T_{00}\rangle_{\chi_m} = -\frac{c}{12m^2}.\label{eq:expect_T}
\end{equation}
Because $\hat{\rho}_A$ is a convex combination of single string reduced density matrices $\hat{\chi}_{m,A}$, the expectation value \eqref{eq:expect_T} can be used to compute the expectation value of the modular Hamiltonian. This turns the inequality \eqref{eq:S_upper} into
\begin{equation}
S\LF \hat{\rho}^{(1)}_A\RF - S\LF \hat{\chi}^{(1)}_{\sqrt{N},A}\RF\leq \frac{\pi c}{3\sqrt{N}}\LT 1 - \sum\limits_{m=1}^N \frac{N_m}{m}\RT\int\limits_0^\ell d\theta \frac{\sin\LF \frac{\ell -\theta}{2\sqrt{N}}\RF\sin\LF \frac{\theta}{2\sqrt{N}}\RF}{\sin\LF \frac{\ell}{2\sqrt{N}}\RF}.\label{eq:S_upper2}
\end{equation}
We define the variable $x=m\sqrt{2\pi/N}$ which is $1/\sqrt{N}$ spaced. This becomes a continuous variable if $N$ is large enough, and the sum is converted into an integral. The integral is dominated by the small $x$ behaviour of the integrand, and is estimated by
\begin{equation}
\sum\limits_{m=1}^N \frac{8}{m\sinh\LF \sqrt{\frac{2\pi}{N}}m\RF} = \int\limits_{\sqrt{\frac{2\pi}{N}}}^{\sqrt{2\pi N}} \frac{8dx}{x\sinh x}\approx 8\sqrt{\frac{N}{2\pi}}.
\end{equation}
The integral in \eqref{eq:S_upper2} can easily be computed and equals
\begin{equation}
\int\limits_0^\ell d\theta \frac{\sin\LF \frac{\ell -\theta}{2\sqrt{N}}\RF\sin\LF \frac{\theta}{2\sqrt{N}}\RF}{\sin\LF \frac{\ell}{2\sqrt{N}}\RF} = \LT \sqrt{N} - \frac{\ell}{2\tan\LF\frac{\ell}{2\sqrt{N}}\RF}\RT \approx \frac{\ell^2}{12\sqrt{N}}.
\end{equation}
Combining all terms we find that the upper bound \eqref{eq:S_upper2} is proportional to $1/\sqrt{N}$ and together with the lower bound \eqref{eq:S_lower} we find that to leading order in the large $N$ limit the entanglement entropy of $\hat{\rho}_A$ is that of an interval on a single $\sqrt{N}$-wound string in the vacuum,
\begin{equation}
S\LF \hat{\rho}^{(1)}_A\RF \approx S\LF \hat{\chi}^{(1)}_{\sqrt{N},A}\RF = \frac{c}{3}\log\LF\frac{\ell}{\epsilon}\RF.
\end{equation} 
The entanglement entropy shows a functional dependence on $\ell/\epsilon$ that one expects from the RT formula \cite{Ryu:2006bv} in a massless BTZ background, but $c$ in our formula is the central charge of the seed CFT instead of the orbifold central charge. This is expected since we have computed the entanglement entropy of a single strand on a spatial interval, not the entanglement entropy of all strands on the spatial interval. The single strand entanglement entropy also agrees with the single strand entwinement as defined in \cite{Balasubramanian:2016xho}. In particular, the single strand entanglement entropy does not have a transition as the size of the interval continues from $\ell<\pi$ to $\ell>\pi$. In the massless BTZ geometry the spatial entanglement entropy does have a transition. For $\ell<\pi$ it is dominated by a minimal geodesic, while for $\ell>\pi$ it is dominated by a disconnected configuration consisting of a minimal geodesic and a surface that wraps the horizon with vanishing area.

\subsubsection{A comment on R\'enyi entropy and replica symmetry breaking}
Suppose we wish to compute the R\'enyi entropy 
\begin{equation}
S^{(n)}\LF\hat{\rho}_A\RF = \frac{1}{1-n}\log\LF\Tr\LF\hat{\rho}^n_A\RF\RF.
\end{equation}
Because $\hat{\rho}_A$ is a convex combination of the density matrices on a single multiwound string, the trace can be expanded into 
\begin{equation}
\Tr\LF\hat{\rho}^n_A\RF = \frac{1}{N^n}\sum\limits_{(m_1,\ldots,m_n)} m_1N_{m_1}\ldots m_nN_{m_n}\Tr\LF\hat{\chi}_{A,m_1}\ldots\hat{\chi}_{A,m_n}\RF.
\end{equation}
The R\'enyi entropies contain cross terms, where not all $m_i$ are equal. These terms can be represented as path integrals over manifolds that are not replica symmetric, namely $n$ cylinders with different radii $m_1,\ldots,m_n$ sewn together. It is an interesting open problem to compute path integrals on such genus zero manifolds.\footnote{Perturbatively in the subsystem size one can use ``cutting and sewing" techniques to calculate this, such as in \cite{Cardy:2014rqa,Mandal:2015jla,Basu:2017kzo}.}

\subsection{Multiple strands}\label{sec:EE_multiple}
 \subsubsection{Conical defects: reduced density matrix and entanglement entropies}
Our formalism of integrating out strands in principle also allows us to compute the entanglement entropy of multiple strands. As an example we will show how it works in the case of the entropy of two strands in the conical defect state, where we integrate out all but a union of a complete strand and an interval $A$ of another strand. \\

As we have argued in sec.~\ref{sec:rhoA_EE_general}, the overlap between terms in the wavefunction and its conjugate that are non-trivially permuted compared to one another, vanishes. The two strand reduced density matrix thus consists of two kinds of overlaps: either the two strands belong to the same $m$-wound string, or they do not. It has the structure\footnote{To explain the normalization factors, we first note that $(N-2)!$ terms contribute equally in \eqref{eq:rhoA_cinC_h} for $l=2$. Second, there are $N$ possible embeddings of $X_{1}$ and $X'_{1}$. Third, either $X_{2,A}$ and $X'_{2,A}$ belong to the same string as $X_{1}$ and $X'_{1}$ or they don't. In the latter case there are $(N-m)$ possible embeddings of $X_{2,A}$ and $X'_{2,A}$ that all contribute equally to $\rho_A^{(2)}$.} 
\begin{align}
\rho_{A}^{(2)} &\LF X_1,X_{2,A}; X'_1,X'_{2,A}\RF = N(N-2)!\left|C_h\right|\LT \sum\limits_{q=2}^m \int  DY_{2,\bar{A}}DY_3\ldots DY_m \right.\\
&\left. \times \psi_m\LF X_1,Y_3,\ldots,X_{2,A},Y_{2,\bar{A}},Y_{q+1},\ldots,Y_m\RF \psi_m^{*}\LF X'_1,Y_3,\ldots,X'_{2,A},Y_{2,\bar{A}},Y_{q+1},\ldots,Y_m\RF\right.\nonumber\\
&\left. + (N-m) \int  DY_3\ldots DY_{m+1} \psi_m\LF X_1,Y_3,\ldots,Y_{m+1}\RF\psi^{*}_m\LF X'_1,Y_3,\ldots,Y_{m+1}\RF\right.\nonumber\\
& \left. \times \int DY_{2,\bar{A}}DY_3\ldots DY_{m+1} \psi_m\LF X_{2,A},Y_{2,\bar{A}},Y_3,\ldots,Y_{m+1}\RF\psi^{*}_m\LF X'_{2,A},Y_{2,\bar{A}},Y_3,\ldots,Y_{m+1}\RF\RT.\nonumber
\end{align}
To make the structure of such a density matrix clearer we can write it schematically as
\begin{equation}
\rho_{A}^{(2)}\LF X_1,X_2;X'_1,X'_2\RF  \propto 
(N-m)\chi^{(1)}_{A,m}(X_1,X'_1)\otimes \chi^{(1)}_{A,m} (X_2,X'_2)
+
 \sum\limits_{q=2}^m \chi^{(2),q}_{A,m}\LF X_1,X_2;X'_1,X'_2\RF ,
\end{equation}
where $q$ denotes how many strands $X_2$ is shifted compared to $X_1$ inside the long string. All of the reduced density matrices $\chi$ involve single multiwound strings of length $m$, so as long as $m$ is smaller than $O(N)$, $\rho_A^{(2)}$ will be dominated by the configurations $\chi$ where $X_1$ and $X_2$ belong to different strings. 
To leading order in $N$ the two strand entanglement entropy of the properly normalized density matrix $\hat{\rho}_A^{(2)}$  is the sum of single string vacuum entropies, namely
\begin{align}
S\LF \hat{\rho}_{A}^{(2)}\RF &=  S\LF \hat{\chi}^{(1)}_{m}\RF + S\LF \hat{\chi}^{(1)}_{A,m}\RF + O\LF \frac{1}{N}\RF,\\
&= \frac{c}{3}\log\LT \frac{2m}{\epsilon}\sin\LF \frac{\pi}{m}\RF\RT +  \frac{c}{3}\log\LT \frac{2m}{\epsilon}\sin\LF \frac{\ell}{2m}\RF\RT  +O\LF \frac{1}{N}\RF.
\end{align}

Notice in particular that the two strand entanglement entropy will not be equal to the entwinement of a long strand. Entwinement is expected to be related to the entanglement entropy of continuously connected strings. Here we have in no sense imposed continuity between the strands in the reduced density matrix, so we do not expect agreement with entwinement.

\section{Discussion and outlook}\label{sec:discussion}
Entanglement entropy of spatial subregions has proven to be a central quantity in the study of holography. The RT formula for spatial entanglement entropy has led to a wealth of discoveries concerning the nature of the holographic dictionary at least up to scales of the order of the AdS radius. To study holography at scales below the AdS radius, it has been argued that one needs to know about the internal degrees of freedom, which are typically gauged \cite{Susskind:1998dq,Susskind:1998vk}.  Hence it is natural  to ask what the entanglement entropy of gauged internal degrees of freedom is. Moreover, since entanglement entropy in field theory is based on a bipartite splitting of the Hilbert space, rather than on a splitting of physical space itself, it is just as natural to study the entanglement entropy of internal degrees of freedom as it is to study the entropy of spatial subregions.

In this paper we studied a symmetric product orbifold CFT. Such a CFT describes the low energy limit of the D1-D5 system and thus appears naturally in the context of holography. States of a symmetric product orbifold describe a collection of multiwound strings, of which the elementary  strands are indistinguishable. Physical states therefore necessarily have to be invariant under permutations. This is reminiscent of a system of identical particles.  
 The entanglement entropy of $k$ out of $N$ internal gauged degrees of freedom in field theory  is then analogous to the entropy of $k$ out of $N$ identical particles.  We showed that the reduced density matrix on $k$ particles does not have support on a subalgebra of operators, but rather on a linear subspace of operators. 

We presented a formula for the reduced density matrix for a general splitting of the internal degrees of freedom of an orbifold CFT into two subsets, and worked out the associated von Neumann entropy for two specific states that are dual in the context of the D1-D5 system to  conical defect geometries and to massless BTZ black holes.
In both cases we find that the entropy of a single strand is computed in the dual geometry by the length of a geodesic.  When the strand has spatial support on an interval of angular extent $\ell<\pi$ it reproduces the length of a minimal geodesic. As such it has the same functional dependence on $\ell$ as the spatial entanglement entropy, but in contrast with the spatial entropy it does not scale with the full central charge of the orbifold theory but rather with the central charge of the seed CFT. This makes sense because the spatial entropy involves having access to all strands on a spatial interval of size $\ell$, in contrast with the single strand entropy. When $\pi<\ell<2\pi$ the single-strand entanglement that we compute is proportional to the length of a non-minimal but non-winding geodesic. It agrees with the single strand entwinement. 

We also studied the entropy of two strands in the conical defect state, and showed that in the large $N$ limit it does not agree with the two strand entwinement. Instead it equals the sum of the entwinement of a single strand with support on the full circle and the single strand entwinement on an interval of size $\ell$. This is because the reduced density matrix is dominated by contributions where the two strands are located on different strings.  Whereas the entropy that we have studied in this paper quantifies the entanglement of several strands with the rest of the system, we expect entwinement to quantify the entanglement of several {\it continuously connected} strands with the rest of the system. To prove that this is really the case, we need a way to  specify continuity across the strands. This would reflect the replica definition of entwinement \cite{Balasubramanian:2016xho}, where one starts with two elementary twist operators on the same strand, averages over $S_N$ permutations, and then moves one of the twist operators to adjacent strands using continuity.   In the construction of the present paper, an interval may span multiple strands that are separately permuted by the gauge group $S_N$. In order to make contact with entwinement the challenge is to define a reduced density matrix associated to continuously connected strands in a way that is manifestly gauge invariant.

We applied our method to the D1-D5 system, but it can also be used for entanglement entropy in matrix string theory, which is also a symmetric product orbifold. Permutation symmetry appears there as the Weyl group of $U(N)$.

\section*{Acknowledgements}
We would like to thank Alex Belin, Alice Bernamonti, William Donnelly, Federico Galli, Arjun Kar, Aitor Lewkowycz and Onkar Parrikar for useful discussions.
This research was supported in part by FWO-Vlaanderen (projects G044016N and G006918N),  by the Vrije Universiteit Brussel through the Strategic Research Program ``High-Energy Physics'', by the Simons Foundation (385592, VB) through the It From Qubit Simons Collaboration, and by the US DOE through Grant FG02-05ER-41367.  Work on this project at the Aspen Center for Physics was supported by NSF grant PHY-1607611. TDJ is aspirant FWO.

\bibliographystyle{jhep}
\bibliography{draft_EE_Tim}

\end{document}